

\font\titlefont = cmr10 scaled\magstep 4
 2
\font\sectionfont = cmr10
\font\littlefont = cmr5 
\font\eightrm = cmr8

\def\ss{\scriptstyle}
\def\sss{\scriptscriptstyle}

\newcount\tcflag
\tcflag = 0  

\ifnum\tcflag = 0 \magnification = 1200 \fi  

\global\baselineskip = 1.2\baselineskip 
\global\parskip = 4pt plus 0.3pt 
\global\abovedisplayskip = 18pt plus3pt minus9pt
\global\belowdisplayskip = 18pt plus3pt minus9pt
\global\abovedisplayshortskip = 6pt plus3pt
\global\belowdisplayshortskip = 6pt plus3pt

\def\barsoff{\overfullrule=0pt}


\def\endignore{}
\def\ignore #1\endignore{} 

\newcount\dflag
\dflag = 0


\def\monthname{\ifcase\month 
\or January \or February \or March \or April \or May \or June%
\or July \or August \or September \or October \or November %
\or December 
\fi}

\newcount\dummy
\newcount\minute  
\newcount\hour
\newcount\localtime
\newcount\localday
\localtime = \time
\localday = \day

\def\advanceclock#1#2{ 
\dummy = #1
\multiply\dummy by 60
\advance\dummy by #2
\advance\localtime by \dummy
\ifnum\localtime > 1440 
\advance\localtime by -1440
\advance\localday by 1
\fi}

\def\settime{{\dummy = \localtime %
\divide\dummy by 60%
\hour = \dummy 
\minute = \localtime%
\multiply\dummy by 60%
\advance\minute by -\dummy 
\ifnum\minute < 10
\xdef\spacer{0} 
\else \xdef\spacer{}
\fi %
\ifnum\hour < 12
\xdef\ampm{a.m.} 
\else
\xdef\ampm{p.m.} 
\advance\hour by -12 %
\fi %
\ifnum\hour = 0 \hour = 12 \fi 
\xdef\timestring{\number\hour : \spacer \number\minute%
\thinspace \ampm}}}



\def\endtitle{}
\def\title#1\endtitle{\vskip.5in\titlefont
\global\baselineskip = 2\baselineskip 
#1\vskip.4in
\baselineskip = 0.5\baselineskip\rm}

\def\endauthors{}
\def\authors#1\endauthors{#1}

\def\endabstract{}
\def\abstract#1\endabstract{\vskip .3in%
\centerline{\sectionfont\bf Abstract}%
\vskip .1in
\noindent#1}

\def\nopageonenumber{\footline={\ifnum\pageno<2\hfil\else
\hss\tenrm\folio\hss\fi}}  

\newcount\nsection 
\newcount\nsubsection 

\def\section#1{\global\advance\nsection by 1
\nsubsection=0
\bigskip\noindent\centerline{\sectionfont \bf \number\nsection.\ #1}
\bigskip\rm\nobreak}

\def\subsection#1{\global\advance\nsubsection by 1
\bigskip\noindent\sectionfont \sl \number\nsection.\number\nsubsection)\
#1\bigskip\rm\nobreak}

\def\topic #1{{\medskip\noindent $\bullet$ \it #1:}}

\def\appendix#1#2{\bigskip\noindent%
\centerline{\sectionfont \bf Appendix #1.\ #2} 
\bigskip\rm\nobreak} 


\newcount\nref 
\global\nref = 1 

\def\therefs{}


\def\ref#1#2{\xdef #1{[\number\nref]} 
\ifnum\nref = 1\global\xdef\therefs{\item{[\number\nref]} #2\ } 
\else
\global\xdef\oldrefs{\therefs}
\global\xdef\therefs{\oldrefs\vskip.1in\item{[\number\nref]} #2\ }%
\fi%
\global\advance\nref by 1
}

\def\listrefs{\vfill\eject\section{References}\therefs}


\newcount\nfoot 
\global\nfoot = 1 

\def\foot#1#2{\xdef #1{(\number\nfoot)} 
\hskip -0.2cm ${}^{\number\nfoot}$
\footnote{}{\vbox{\baselineskip=10pt
\eightrm \hskip -1cm ${}^{\number\nfoot}$ #2}}
\global\advance\nfoot by 1
}


\newcount\nfig 
\global\nfig = 1
\def\thefigs{} 

\def\figure#1#2{\xdef #1{(\number\nfig)}
\ifnum\nfig = 1\global\xdef\thefigs{\item{(\number\nfig)} #2\ }
\else
\global\xdef\oldfigs{\thefigs}
\global\xdef\thefigs{\oldfigs\vskip.1in\item{(\number\nfig)} #2\ }%
\fi%
\global\advance\nfig by 1 } 

\def\fig#1{\xdef #1{(\number\nfig)}
\global\advance\nfig by 1 } 


\newcount\ntab
\global\ntab = 1

\def\table#1{\xdef #1{\number\ntab}
\global\advance\ntab by 1 } 


\newcount\cflag
\newcount\nequation
\global\nequation = 1
\def\eqlabel{(1)}

\def\nexteqno{\ifnum\cflag = 0
\global\advance\nequation by 1
\fi
\global\cflag = 0
\xdef\eqlabel{(\number\nequation)}}

\def\lasteqno{\global\advance\nequation by -1
\xdef\eqlabel{(\number\nequation)}}

\def\label#1{\xdef #1{(\number\nequation)}
\ifnum\dflag = 1
{\escapechar = -1
\xdef\draftname{\littlefont\string#1}}
\fi}

\def\clabel#1#2{\xdef\eqlabel{(\number\nequation #2)}
\global\cflag = 1
\xdef #1{\eqlabel}
\ifnum\dflag = 1
{\escapechar = -1
\xdef\draftname{\string#1}}
\fi}

\def\cclabel#1#2{\xdef\eqlabel{#2)}
\global\cflag = 1
\xdef #1{\eqlabel}
\ifnum\dflag = 1
{\escapechar = -1
\xdef\draftname{\string#1}}
\fi}


\def\eeq{}

\def\eqnn #1\eeq{$$ #1 $$}

\def\eq #1\eeq{
\ifnum\dflag = 0
{\xdef\draftname{\ }}
\fi 
$$ #1
\eqno{\eqlabel \rlap{\ \draftname}} $$
\nexteqno}







\def\eqa #1\eeq{
\ifnum\dflag = 0
{\xdef\draftname{\ }}
\fi 
$$ \eqalignno{ #1 } $$
\global\cflag = 0}


\def\ie{{\it i.e.\/}}
\def\eg{{\it e.g.\/}}

\def\via{{\it via\/}}


\def\anp#1#2#3{{\it Ann.\ Phys. (NY)} {\bf #1} (19#2) #3}

\def\npb#1#2#3{{\it Nucl.\ Phys.} {\bf B#1} (19#2) #3}
\def\plb#1#2#3{{\it Phys.\ Lett.} {\bf #1B} (19#2) #3}

\def\prd#1#2#3{{\it Phys.\ Rev.} {\bf D#1} (19#2) #3}

\def\prl#1#2#3{{\it Phys.\ Rev.\ Lett.} {\bf #1} (19#2) #3}


\global\nulldelimiterspace = 0pt



\def\frac#1#2{{{#1} \over {#2}}\,}  
\def\hf{{1\over 2}}
\def\nth#1{{1\over #1}}


\def\Asl{\hbox{/\kern-.7500em\it A}} 
\def\Dsl{\hbox{/\kern-.6700em\it D}} 
\def\dsl{\hbox{/\kern-.5300em$\partial$}}
\def\pxpsl{\hbox{/\kern-.5600em$p$}}
\def\sslsh{\hbox{/\kern-.5300em$s$}}
\def\epssl{\hbox{/\kern-.5100em$\epsilon$}}
\def\delsl{\hbox{/\kern-.6300em$\nabla$}}
\def\lxpsl{\hbox{/\kern-.4300em$l$}}
\def\elxpsl{\hbox{/\kern-.4500em$\ell$}}
\def\kxpsl{\hbox{/\kern-.5100em$k$}}
\def\qxpsl{\hbox{/\kern-.5000em$q$}}
\def\sla#1{\raise.15ex\hbox{$/$}\kern-.57em #1}



\def\roughly#1{\mathrel{\raise.3ex\hbox{$#1$\kern-.75em\lower1ex\hbox{$\sim$}}}}

\def\ol#1{\overline{#1}}





\def\Scw{{\cal W}}


\def\ssa{{\sss A}}
\def\ssb{{\sss B}}

\def\ssl{{\sss L}}

\def\ssr{{\sss R}}
\def\ssS{{\sss S}}
\def\sst{{\sss T}}

\def\ssw{{\sss W}}


\def\pmb#1{\setbox0=\hbox{#1}%
\kern-.025em\copy0\kern-\wd0
\kern.05em\copy0\kern-\wd0
\kern-.025em\raise.0433em\box0}


\font\jlgtenbrm=cmbx10
\font\jlgtenbit=cmmib10
\font\jlgtenbsy=cmbsy10
\font\jlgsevenbrm=cmbx10 at 7pt
\font\jlgsevenbsy=cmbsy10 at 7pt
\font\jlgsevenbit=cmmib10 at 7pt
\font\jlgfivebrm=cmbx10 at 5pt
\font\jlgfivebsy=cmbsy10 at 5pt
\font\jlgfivebit=cmmib10 at 5pt
\newfam\jlgbrm

\textfont\jlgbrm=\jlgtenbrm
\scriptfont\jlgbrm=\jlgsevenbrm
\scriptscriptfont\jlgbrm=\jlgfivebrm
\newfam\jlgbit

\textfont\jlgbit=\jlgtenbit
\scriptfont\jlgbit=\jlgsevenbit
\scriptscriptfont\jlgbit=\jlgfivebit
\newfam\jlgbsy

\textfont\jlgbsy=\jlgtenbsy
\scriptfont\jlgbsy=\jlgsevenbsy
\scriptscriptfont\jlgbsy=\jlgfivebsy
\newcount\jlgcode
\newcount\jlgfam
\newcount\jlgchar
\newcount\jlgtmp
\def\bolded#1{
        \jlgcode\the#1 \divide\jlgcode by 4096
        \jlgtmp\the\jlgcode \multiply\jlgtmp by 4096
        \jlgfam\the#1 \advance\jlgfam by -\the\jlgtmp
        \divide\jlgfam by 256
        \jlgtmp\the\jlgcode \multiply\jlgtmp by 16
	\advance\jlgtmp by \the\jlgfam
	\multiply\jlgtmp by 256
        \jlgchar\the#1 \advance\jlgchar by -\the\jlgtmp
        \advance\jlgfam by \the\jlgbrm
        \jlgtmp\the\jlgcode
        \multiply\jlgtmp by 16
        \advance\jlgtmp by \the\jlgfam
        \multiply\jlgtmp by 256
        \advance\jlgtmp by \the\jlgchar
        \mathchar\the\jlgtmp
}


\def\tr{\mathop{\rm tr}}
\def\Tr{\mathop{\rm Tr}}
\def\det{\mathop{\rm det}}

\def\Re{{\rm Re\;}}



\def\Avg#1{\left\langle #1 \right\rangle}






\nopageonenumber
\baselineskip = 18pt
\barsoff

\def\bk{\item{}}

\def\IR{\relax{\rm I\kern-.18em R}}
\font\cmss=cmss10 \font\cmsss=cmss10 at 7pt
\def\IZ{\relax\ifmmode\mathchoice
{\hbox{\cmss Z\kern-.4em Z}}{\hbox{\cmss Z\kern-.4em Z}}
{\lower.9pt\hbox{\cmsss Z\kern-.4em Z}}
{\lower1.2pt\hbox{\cmsss Z\kern-.4em Z}}\else{\cmss Z\kern-.4em Z}\fi}


\line{hep-th/9707062 \hfil McGill-97/07, IFUNAM  FT97-10}

\vskip .1in
\title
\centerline{Fixing the Dilaton with}
\centerline{Asymptotically-Expensive Physics?}
\endtitle

\vskip 0.2in
\authors
\centerline{C.P. Burgess${}^a$, A. de la Macorra${}^b$, I. Maksymyk${}^c$ and
F. Quevedo${}^b$}
\vskip .1in
\centerline{\it ${}^a$ Physics Department, McGill University}
\centerline{\it 3600 University St., Montr\'eal, Qu\'ebec, Canada, H3A 2T8.}
\vskip .05in
\centerline{\it ${}^b$ Instituto de F\'isica, Universidad Nacional  Aut\'onoma
de M\'exico}
\centerline{\it Apartado Postal 20-364, 01000  M\'exico D.F., M\'exico.}
\vskip .05in
\centerline{\it ${}^c$ TRIUMF, 4004 Wesbrook Mall}
\centerline{\it Vancouver, British Columbia, Canada, V6T 2A3.}
\endauthors

\abstract
\vbox{\baselineskip 15pt
 We propose a general mechanism for stabilizing the dilaton against runaway to
weak coupling. The method is based on features of the effective superpotential
which arise for supersymmetric gauge theories which are not asymptotically
free. Consideration of the 2PI effective action for bilinear operators of
matter and gauge superfields allows one to overcome the obstacles
to constructing a nonvanishing superpotential. }
\endabstract


\vfill\eject

\section{Introduction}

String theorists would love to  calculate the mass of
the electron from first principles. Indeed, a quantitative determination
of the elementary-particle spectrum predicted for string models
at the weak scale is a prerequisite for any convincing comparison
between string theory and experiment. The main obstacle to
making such a determination has been a lack of understanding
of how supersymmetry is spontaneously broken in string models,
since supersymmetry breaking ultimately controls the
pattern of masses which would be observed at experimentally-accessible
energies.

\ref\dynamicalbreaking{E. Witten, \npb{188}{81}{513};
\npb{202}{82}{253}.}
\ref\gcondsn{J.-P. Derendinger, L.E. Ib\'a\~nez, H.P. Nilles,
\plb{155}{85}{65}; M. Dine, R. Rohm, N.Seiberg and E. Witten,
\plb{156}{85}{55}.}
\ref\AKMRV{For a review, see D. Amati, K. Konishi, Y. Meurice,
G. C. Rossi and G. Veneziano, {\it Phys. Rep.} {\bf 162} (1988) 169.}

\ref\factors{N.   Krasnikov, \plb{193}{87}{37}.}

Our present inability to understand supersymmetry breaking in string
theory is not for want of trying. It was among the
first problems addressed by early workers, who discovered the
tantalizingly attractive mechanism for dynamical supersymmetry breaking
\dynamicalbreaking\ \via\ gaugino condensation \gcondsn, \AKMRV.
Unfortunately, however promising its beginnings, this mechanism
has proven to have some difficulties. One of these is
its prediction of a superpotential which is exponential in the dilaton
supermultiplet, $W = A \; e^{- a \, S}$, with $a$ a positive constant.
The resulting scalar potential is minimized for $\Re \, S \to
\infty$, in which limit the gauge coupling vanishes and
supersymmetry does not break.

\ref\rami{R. Brustein and P. Steinhardt, \plb{302}{93}{196}.}

Generically, the scalar potential may have other minima in addition
to the one at infinity. This happens, for example, when gaugino
condensation occurs for a gauge group consisting of several
factors \factors, in which case the resulting superpotential
has the form of a sum of exponentials:
\label\expsum
\eq
W = \sum_n A_n \; e^{- a_n \, S}.
\eeq
The problem with this scenario is that it is difficult to understand why the
universe should not end up within the basin of attraction of
the `runaway' solution  \rami.

\ref\genericprob{M. Dine and N. Seiberg, \plb{162}{85}{299}.}

It has been argued that the `runaway dilaton'
problem is generic in string theory \genericprob\ .
This is because the {\it v.e.v.}s of
the dilaton multiplet's complex scalars determine the coupling
constant, $g$, and vacuum angle, $\theta$, of the low-energy
gauge interactions according to the relation $\Avg{S} =
{1 \over g^2} + {i \theta \over 8 \pi^2}$. Since flat space is known
to solve the field equations for noninteracting strings --- \ie\ for
$g=0$ --- the scalar potential might be expected to always
admit a minimum when $\Re \, S \to \infty$.

The purpose of this letter is to point out some loopholes in the
argument that the superpotential must vanish for large $S$. We do
so by constructing some simple models which do not have this
property.  One class of models to which we are led involves low-energy
gauge groups which are {\it not} asymptotically free. We therefore
devote some discussion to special issues which arise when
using the effective superpotentials for nonasymptotically-free
effective gauge theories.

\ref\filq{A. Font, L. Ib\'a\~nez, D. L\"ust  and F. Quevedo,
\plb{249}{90}{35}.}

We start by re-examining the superpotential, eq.~\expsum,
arising from traditional gaugino condensation. Our main
observation is that the problem of runaway solutions only arises
if all of the exponents in this equation are negative.
For instance, in global supersymmetry
the scalar potential is minimized by the extrema of the
superpotential, so if both $a_1$ and $a_2$ are positive, then:
\label\newexpsum
\eq
W = A_1 \; e^{- a_1 \, S} + A_2 \; e^{+ a_2 \, S} ,
\eeq
has only the single vacuum solution:\foot\sdual{Dilaton
potentials blowing up at infinity were discussed in
 the context of $\ss S$ duality in \filq.}
$\ol{S} = \ln\left( a_1 A_1/a_2 A_2 \right) /
(a_1 + a_2)$.

\ref\classic{E. Witten, \plb{155}{85}{151}; \bk
C.P. Burgess, A. Font and F. Quevedo, \npb{272}{86}{661}.}

\ref\noscale{E. Cremmer, S. Ferrara, C. Kounnas and D.V. Nanopoulos,
\plb{133}{83}{61}; J. Ellis, C. Kounnas and D.V. Nanopoulos,
\npb{241}{84}{406}, \npb{247}{84}{373}.}

\ref\Tpotentials{For a review with  references
 see  F. Quevedo
hep-th/9603074.}

Once coupled to supergravity this solution can also break
supersymmetry, depending on the form taken by the
K\"ahler potential, $K$. For example, the usual perturbative
string theory result \classic: $K = - \ln(S+S^*)
- 3 \ln(T + T^*)$ --- where $T$ denotes the complex scalar
containing the `breathing'
modulus --- leads to the `no-scale' \noscale\ scalar
potential $V = |D_\ssS W|^2 (S+S^*)/(T+T^*)^3$. Here
$D_i W \equiv \partial_i W + \partial_i K \; W$ (with
$i = S,T$) are order parameters for supersymmetry breaking.
$V$ is minimized by ensuring $D_\ssS W = 0$, for any
$T$. So long as other contributions to the potential
do not drive $T + T^* \to \infty$,
$D_\sst W \ne 0$ at the minimum of $V$, and so supersymmetry
is broken.
A similar thing happens if the theory is required to be invariant
under $T$ duality. In this case $W$ depends nontrivially on $T$,
although $T$-duality invariance nonetheless ensures that
$ |D_\ssS W|^2=0$ but $ |D_\sst W|^2\not =0$ at the minimum,
 breaking supersymmetry \Tpotentials.

Of course, if the supersymmetry-breaking scale in
the above scenario is taken to be much smaller
than the Planck mass --- which is taken here as unity ---  then
some fine-tuning is required of the parameters $A_i$ or $a_i$.
Here we do not pursue the extent to
which this fine-tuning may be ameliorated, but focus instead
on the runaway-dilaton problem, and how a superpotential like
eq.~\newexpsum\ might be generated in the first place.

\section{Non-Asymptotically Free Models}

\ref\bdqq{C.P. Burgess, J.-P. Derendinger, F. Quevedo and
M. Quiros, \anp{250}{96}{193}.}

\ref\KL{V. Kaplunovsky and J. Louis, \npb{422}{94}{57}.}

\ref\compfp{See, \eg, P.M. Chaikin and T.C. Lubensky,
{\it Principles of Condensed Matter Physics}, Cambridge
University Press, 1995.}

Some intuition as to how to generate a superpotential like
eq.~\newexpsum, which blows up for small couplings,
can be obtained as follows. Consider a quartic
potential for a real scalar field: $V(\phi) =
- \, \hf \, m^2 \phi^2 + \nth4 \, g^2 \, \phi^4$.
When evaluated at its minimum, $\ol\phi = \pm m/g$,
the scalar potential becomes $V(\ol\phi) =
- m^4/4g^2$, a result which blows up as $g \to 0$.
The potential, once minimized, is singular in the
weak-coupling limit because $g$ premultiplies the
highest power of $\phi$ in $V(\phi)$, and so any
nonzero value for $g$, no matter how small, qualitatively
alters the field configuration which minimizes $V$.
Operators, such as $\nth4 \, g^2 \, \phi^4$, whose
couplings can appear singularly in this way are
well known in renormalization-group applications
within condensed-matter physics, where they can play
important roles in analyses even if they are nominally
irrelevant in the RG sense. They are known there
as `dangerous irrelevant operators' \compfp.
This suggests that an exponential of $+S$
might be achieved by introducing a field whose
highest-dimension term in $W$ arises premultiplied by
a positive power of $e^{-S}$.

Another useful piece of intuition comes from
the observation that strongly-coupled
supersymmetric gauge theories tend to produce
superpotentials which are proportional to the
appropriate power of the renormalization-group invariant
scale: $W \propto \Lambda^3$, where for pure
gauge theories $\Lambda \propto M_s
e^{-8\pi^2/b g^2}$, where $M_s$ is a high-energy
cutoff (the string scale, say) and $b/16\pi^2$
is the coefficient
of the one-loop beta function.\foot\notechange{Our
normalization of $\ss b$ here differs by a factor of
$\ss 16 \pi^2$ from our earlier conventions \bdqq.}
Since it is the gauge coupling at the string scale
which is related to the dilaton by, $1/g^2(M_s) \propto
\Re \, S$, the negative argument of the exponential
is related to the sign of the coefficient, $b$.
This suggests\foot\butotoh{Of course this argument is
merely suggestive, since in the presence
of matter it is the conformal anomaly coefficient, $\ss c$,
rather than the beta-function coefficient, $\ss b$, which
appears in the exponent with $\ss S$ in $\ss W$ \KL.}
 the potential utility of working with
nonasymptotically free (NAF) gauge
theories.

\ref\banks{T. Banks and M.Dine, \prd{50}{94}{7454};
P. Binetruy, M.K. Gaillard and Y.-Y. Wu, \npb{481}{96}{109};
J.A. Casas, \plb{384}{96}{103}.}

Based on these observations, we attempt to obtain
eq.~\newexpsum\ by constructing a model for which
the gauge group consists of two factors, $G = G_1
\times G_2$, with $G_1$ asymptotically free and $G_2$
not. In this case, there is no value of $Re \, S$
for which {\it both} $\Lambda_1$ and $\Lambda_2 \to 0$,
and so for which $W$ might vanish.
For simplicity, we take $G_1$ to consist of a gauge theory
without matter, and focus in what follows on the
physics associated with the second factor,
$G_2$.\foot\banksref{In general we will not need to have two gauge groups
to stabilize the dilaton, we may in principle use corrections
to the K\"ahler potential together with one single exponential
\banks.}

\ref\seiberg{For a review see K. Intriligator and N. Seiberg,
{\it Nucl. Phys. B (Proc. Suppl.) 45B,C (1996) 1}.}

\ref\peskin{M. Peskin, hep-th/9702094.}

\ref\shifman{M. Shifman, hep-th/9704114.}

To make this precise consider supersymmetric Quantum
Chromodynamics (SQCD) having $N_c$ colours and $N_f$
flavours. The superfields of the theory consist of a
left-chiral, left-handed-spinor gauge supermultiplet,
$\Scw^a$ ($a=1,\dots,N_c^2 -1$), and left-chiral scalars
representing the quarks and antiquarks: ${Q^i}_\alpha,
{\ol{Q}_i}^\alpha$ ($i=1,\dots,N_f$, $\alpha = 1,\dots,N_c$).
In the absence of a superpotential the global internal
symmetries of the model
(modulo anomalies) are $G_f = SU_\ssl(N_f) \times
SU_\ssr(N_f) \times U_\ssb(1) \times U_\ssa(1) \times
U_\ssr(1)$, with respect to which the fields are assigned
the following transformation rule:
\label\trules
\eq
\Scw^a \sim \left( {\bf 1}, {\bf 1}, 0, 0, {3 \over 2} \right),
\qquad {Q^i}_\alpha \sim \Bigl( {\bf N_f}, {\bf 1}, 1, 1, 1 \Bigr),
\qquad {\ol{Q}_i}^\alpha \sim \Bigl( {\bf 1}, {\bf \ol{N}_f}, -1, 1, 1 \Bigr).
\eeq
The first two numbers in these expressions denote the representation
of $SU_\ssl(N_f) \times SU_\ssr(N_f)$ under which the field
transforms, while the last three numbers give their charges under
the three $U(1)$ generators, $B, A$ and $R$.

For SQCD the beta function and scale anomaly coefficients are
$b = 3 N_c - N_f$ and $c = N_c - N_f$. For $N_f < 3 N_c$ this
theory is asymptotically free. The possible  nonperturbative
superpotentials and general structure of the theory
 have been extensively investigated for any value of $N_f$ and $N_c$
\seiberg.\foot\reviews{For recent reviews see refs.~\peskin\
and \shifman.}

\ref\ads{I. Affleck, M. Dine and N. Seiberg, \prl{51}{83}{1026}.}

Based on the $\nth4 \, g^2 \phi^4$ analogy presented above,
we supplement SQCD with an additional colour-singlet
left-chiral scalar superfield, ${\mu^i}_j$, which transforms
in the same way as would a quark mass term: ${\mu^i}_j \sim
\Bigl( {\bf \ol{N}_f}, {\bf N_f}, 0, -2, 1 \Bigl)$. We take the
superpotential in the Higgs phase to be \ads\
\label\spothiggs
\eq
W = \Tr \Bigl( \mu Q \ol{Q} \Bigr) + {\lambda \over 3}
\, \Tr \Bigl( \mu^3 \Bigr).
\eeq
where $\lambda$ is a Yukawa coupling which explicitly breaks the
axial symmetry, $U_\ssa(1)$,  but none of the other symmetries.
In fact a cubic superpotential in $\mu$ is the only possible
form which does not explicitly break the $R$-symmetry.
Furthermore, we imagine $\lambda$ to be small enough to justify
ignoring this coupling in comparison with the gauge couplings
when determining the vacuum structure of the model.

\ref\holomorphy{M.A. Shifman and A.I. Vainshtein, \npb{277}{86}{456};
\npb{359}{91}{571}.}

Since our goal is to determine the potential which fixes the
{\it v.e.v.} of the dilaton field, $S$, we now construct the
effective action, $\Gamma$, which generates the model's irreducible
correlation functions.\foot\notwilson{Notice that $\ss \Gamma$ is
$\ss not$ the Wilson action of the theory. It is therefore
potentially subject to holomorphy anomalies, which we believe
to play no role here, in the presence of massless
particles \holomorphy.} We take the arguments of
$\Gamma$ to be $S$ and ${\mu^i}_j$, as well as the
 expectation values of the gauge-invariant
fields  ${M^i}_j \equiv \Avg{{Q^i}_\alpha \,
{\ol{Q}_j}^\alpha}$ and $U \equiv \Avg{\Scw^a \, \Scw_a}$.
(For $N_c < N_f$ we imagine the expectation value of the
baryon operator, $B^{i_1\cdots i_{N_c}} = \Avg{ \epsilon^{\alpha_1
\cdots \alpha_{N_c}} \, {Q^{i_1}}_{\alpha_1} \cdots
{Q^{i_{N_c}}}_{\alpha_{N_c}}}$  to be zero.)

A great deal is known about the exact superpotential which
appears in $\Gamma[U,M,\mu,S]$. The non-perturbative terms are
required on general grounds to be linear in both
$S$ and ${\mu^i}_j$ \bdqq. Furthermore, its dependence
on $U$ and ${M^i}_j$ is completely determined by the gauge and global
symmetries of the problem \seiberg, together with the anomaly-cancelling
transformation rules for $S$: $e^{- 8 \pi^2 S} \sim \Bigl( {\bf 1}, {\bf1},
0, 2 N_f, 3N_c - N_f \Bigr)$. After eliminating $U=U(M,\mu,S)$ by
extremizing $\Gamma$ with respect to $U$, the result becomes:
\label\Wresult
\eq
W(M,\mu,S) = \Tr \Bigl( \mu M \Bigr)+ {\lambda \over 3}
\, \Tr \Bigl( \mu^3 \Bigr)  + k \; \left( {e^{- 8 \pi^2 S} \over
\det M } \right)^{1 /( N_c - N_f)} ,
\eeq
where $k=N_c-N_f$.
Extremizing with respect to ${M^i}_j$, and substituting
the result back into $\Gamma$ then gives the superpotential
\label\Wresulttwo
\eq
W(\mu,S) = {\lambda \over 3} \, \Tr \Bigl( \mu^3 \Bigr) + k'
\Bigl( e^{- 8 \pi^2 S} \; \det \mu \Bigr)^{1 / N_c},
\eeq
where $k'=N_c$. If ${\mu^i}_j$ were
a constant mass matrix, eq.~\Wresulttwo\ would give the
superpotential for $S$ in SQCD. It is noteworthy that,
so long as $k' \ne 0$ and $\mu$ is held fixed,
the result has runaway behaviour
to $S\to \infty$ {\it regardless} of the values of $N_c$
and $N_f$.

The final step is now to extremize $W$ with respect to the
field ${\mu^i}_j$, to obtain the overall superpotential for
$S$. The extremum is obtained for
${\mu^i}_j=\left(-\lambda\, e^{8\pi^2 S/N_c}
\right)^{N_c/(N_f-3N_c)} {\delta^i}_j$, and using this in eq.~\Wresulttwo\
gives:
\label\WSpot
\eq
W(S) = k'' \Bigl( \lambda^{N_f} \; e^{24 \pi^2 S} \Bigr)^{1 /
(N_f - 3N_c)} ,
\eeq
with $k''=(-1)^{3N_c/(N_f-3N_c)}\left(N_f-3N_c\right)/3$.
Notice that eq.~\WSpot\ takes the simple form $W \propto \Lambda^3$
when expressed in terms of the renormalization group invariant scale,
$\Lambda \propto e^{- 8 \pi^2  S/b}$.
Eq.~\WSpot\ gives the desired positive exponential of $S$, but
only if $N_f > 3 N_c$, and only if the theory is not
asymptotically free. When this is combined with the potential
for another, asymptotically-free, factor of the gauge group
we obtain a superpotential of the form of eq.~\newexpsum.

Several comments bear emphasis at this point.
Notice first that the singular dependence implied by eq.~\WSpot\
for $W$ in the limit $S \to \infty$ is very similar to the
singularity of $V(\ol\phi)$ in $\nth4 \, g^2 \, \phi^4$
theory as $g \to 0$. This is most easily seen by examining
eq.~\Wresulttwo\ for the special case where ${\mu^i}_j$ is
proportional to the unit matrix, ${\mu^i}_j \equiv \mu \,
{\delta^i}_j$. In this case the first term of \Wresulttwo\
is cubic in $\mu$, while the second involves $\mu$ raised
to the $N_f/N_c$ power. The second term is therefore
`dangerous' --- \ie\ involves the higher power of
$\mu$ --- precisely when $N_f > 3 N_c$. Since
a positive power of $e^{-8 \pi^2
S}$ premultiplies this second term, this is also when
$W$ is singular as $S \to \infty$.

The possibility of having potentials for $S$ which
diverge as $S \to \infty$ is reminiscent of what
was found years ago for the $T$ field. In this
case, even though large $T$ corresponds to weak
worldsheet coupling, standard $T$-dual potentials
blow up as $\Re T\to \infty$ \Tpotentials. The underlying
reasons for these divergent behaviours appears to
differ in detail, however, since the large-$T$
singularity can be attributed to a large value of
the 10D string coupling, $g_{10}$, even when the 4-D
coupling, $g_4$, is weak. These limits are mutually
consistent for large $\Re T$ (or, equivalently, large
compactification radius, $R$) because $g_{10}$ and $g_4$ are
related by $1/g_4^2 = R^6/g_{10}^2$.

Finally, since ${\mu^i}_j$ plays the role of a quark mass matrix, its
eigenvalues must be smaller than the scale $\Lambda$ if
there is to be a range of scales for which the theory is to be
weakly coupled and not asymptotically free. We may always
ensure this to be true by taking the Yukawa coupling, $\lambda$,
to be sufficiently small. When this is done, the degrees of
freedom in the energy range $\mu < E < E_{\rm max}$,
where $E_{\rm max} = \Lambda$, or some other scale at which
new degrees of freedom become important, describe
a non-asymptotically-free theory of supersymmetric
quarks and gluons. For $E < \mu$ the quarks and their
superpartners may be integrated out, leaving an asymptotically-free
model at these lower scales.

\section{Discussion}

We see that both signs are possible for the arguments of the
exponentials which appear in the superpotential for $S$,
depending on the matter content of the strong-coupling
physics which generates it. In particular, the desired
positive exponentials of $S$ can be generated by non-asymptotically
free SQCD-like gauge theories.

There are a number of criticisms which might be raised against
using nonasymptotically-free gauge theories in the way we have.
We now address some of these.

\topic{(1)}
Perhaps the easiest objection to deal with is the widely
held belief that only asymptotically-free gauge theories
can be consistently defined as interacting quantum
field theories. Even if this proves to be true,
asymptotic freedom is not a fundamental requirement for
an {\it effective} theory, which only describes the degrees
of freedom at very low energies (or very long distances).
Quantum Electrodynamics is probably the most famous
example of such an effective theory, which is now
believed to be the low-energy approximation to the
more complete Standard Model of electroweak interactions.

This point of view is all the more inevitable within the context
of string theory, where the entire discussion of four-dimensional
supersymmetric theories is intended as a low-energy description
of a more fundamental string theory. Furthermore, many
four-dimensional string compactifications are known which
actually have low-energy spectra giving rise to nonasymptotically-free
gauge interactions within the effective low-energy theory
\Tpotentials.

\topic{(2)}
A potentially more compelling objection argues that, although
an effective superpotential like eq.~\Wresult\ can arise in
SQCD for sufficiently small $N_f$, it does {\it not} arise
when $N_f > 3 N_c$. This line of argument proceeds in
one of two ways.

\ref\shanta{S.P. de Alwis, preprint COLO-HEP-363, 
(hep-th/9508053).}

One form of this objection argues that the quantity
$\det M$ itself vanishes for $N_c < N_f$ because the
same is true for $\det \Bigl( Q \ol{Q} \Bigr)$
\peskin, \shifman.\foot\whyzero{$\ss
\det ( Q \ol{Q} ) = 0$ because $\ss {Q^i}_\alpha {\ol{Q}_j}^\alpha$,
being a sum of dyadics, always has zero eigenvalues so long
as $\ss N_c < N_f$.}
The weakness in this argument lies
in its making an insufficient distinction between the
effective action and the Wilson action.\foot\seeus{See
refs.~\bdqq\ and \shanta\ for a detailed discussion of this distinction
within the context of gaugino condensation.} The Wilson
action, $S_\ssw$, describes the dynamics of the low-energy degrees
of freedom of a given system, and is used in the path
integral over these degrees of freedom in precisely the
same way as is the classical action. The Wilson action for SQCD
at scales for which quarks and gluons are the relevant
degrees of freedom would therefore depend on the
fields $\Scw^a$, ${Q^i}_\alpha$ and ${\ol{Q}_i}^\alpha$.
As a result, the vanishing of $\det ( Q \ol{Q} )$ would
indeed preclude the generation
of  a superpotential of the type $\Bigl[ e^{-8 \pi^2 S}/
\det(Q \ol{Q} ) \Bigr]^{1/(N_c - N_f)}$ within the Wilson action.

By contrast, it is the effective action, $\Gamma$, which is
of interest when computing the {\it v.e.v.}s of various
fields. And it is ${M^i}_j = \Avg{{Q^i}_\alpha
{\ol{Q}_j}^\alpha}$ which appears as an argument of $\Gamma$.
Since the expectation of a product of operators is not
equal to the product of the expectations of each operator,
it need not follow that $\det M = 0$ when $N_c < N_f$ \AKMRV.

\topic{(3)}
It is equally clear that the failure of the fields ${M^i}_j$,
$B^{i_1\cdots i_{N_c}}$ and $\ol{B}_{i_1\cdots i_{N_c}}$ to
satisfy the `t Hooft anomaly-matching conditions \peskin\
does not argue against the superpotential \Wresult. Anomaly
matching says that the physical degrees of freedom at any scale
must have anomalies which reproduce the anomaly of the
underlying degrees of freedom. This is a constraint on which fields
can appear in the path integral over the Wilson action at any scale,
since it is the functional measure for these fields which gives the
anomaly. But it is {\it not} a constraint on the 2PI action which we are
considering since the arguments of this action are not quantum fields
to be integrated over, and so do not have anomalies. Anomalies appear
in the 2PI action simply as terms which explicitly break the
corresponding symmetry, and this has been used to construct
the superpotential of \Wresult.

\topic{(4)}
An alternative objection concedes the necessity of using the
effective action, $\Gamma$, rather than the Wilson action, $S_\ssw$.
It also concedes the consistency of eq.~\Wresult\ with all of
the symmetries of the theory. The objection is that the constant
$k$, which premultiplies the interesting term in this equation,
must equal zero.

The vanishing of $k$ might be argued as follows.
For asymptotically-free theories the last term of
eq.~\Wresult\ is interpreted as being
the result of gaugino condensation.
This may be seen explicitly by recognizing that the
gaugino condensate is given by $U = \partial W/\partial S$,
with $W$ given by eq.~\Wresult, \Wresulttwo\ or \WSpot.
If the theory is not asymptotically free, the argument continues,
then it is weakly coupled at low energies and so the low-energy
dynamics cannot cause the gauginos to condense (or
have any other strong-coupling consequences).

One difficulty with this argument is that it assumes it is the low-energy
dynamics which must be responsible for the nonperturbative
terms in the effective superpotential. In a theory which is not
asymptotically free, it is the higher energy degrees of freedom
which are more strongly coupled than the lower energy ones.
As a result one would expect a term of the form
of eq.~\Wresulttwo\ to be directly generated in the Wilson action
at high energies. This term can then survive down to low energies
to contribute to $\Gamma$.
A second, perhaps more convincing, difficulty with this
argument is given in the next item.

\topic{(5)}
A variation of the previous point argues that instanton
calculations do not produce such a nonvanishing superpotential
\shifman. However, instanton calculations do not always
explicitly produce nonzero contributions to quantities
which are known not to vanish. For instance, instantons
give a vanishing gaugino bilinear, $\Avg{\ol\lambda \lambda}$,
because there are not enough fermion fields to soak up all
of the zero modes which arise in the instanton path integral.
We nevertheless know that the
expectation value of the gaugino bilinear does not vanish
since it can be determined by factorizing the
expectation values of correlations of more powers of
gaugino fields, for which instanton contributions are
nonzero \AKMRV.
Therefore, we regard the absence
of instanton contributions to be insufficient
evidence for a vanishing superpotential.

Perhaps the best argument against this, and the previous, point
is based on continuity with
the case $N_c \ge N_f$ \AKMRV. A powerful technique for determining
the constants, $k$, $k'$ and $k''$ in the superpotentials is to
give large masses to some of the quarks, and then to integrate
these out. Once this is done one must obtain the correct result
for the theory having fewer quarks. This permits relations to be
derived between theories having the same value of $N_c$ but
differing values of $N_f$. If it is supposed that no nonperturbative
superpotential can appear for $N_f > 3 N_c$, then it is difficult
to obtain the known superpotential in the asymptotically-free
case where $N_f$ is smaller. The same objection does not
apply if eq.~\Wresult\ applies for {\it all} values of $N_c$ and
$N_f$.

\ref\seibergdual{N. Seiberg, \npb{435}{95}{129}.}

\topic{(6)}
With our final point we address a minor puzzle.
According to Seiberg, SQCD with $N_f > N_c+1$
is dual to a supersymmetric model having $N_f$
flavours and $\widetilde N_c\equiv N_f-N_c$ colours,
coupled to an extra gauge-singlet field, $\widetilde {M^i}_j$,
which has the quantum numbers of a mass matrix for the dual quarks,
${q}^i$. Since the condition for asymptotic freedom is $N_f < 3 N_c$
for the original theory, and $N_f > 3 N_c/2$ for its dual, it is
impossible that {\it both} the original theory and its dual
are not asymptotically free. The puzzle is this: for
$N_f > 3 N_c$ the dual theory is asymptotically free
and so should dynamically generate a superpotential
which vanishes in the weak-coupling limit. How can this
share the same vacuum structure as the model
constructed using the original variables?

To see that the dual theory indeed implies a superpotential
which vanishes at weak coupling, even though it also
involves the new gauge-singlet field, ${M^i}_j$,
which is otherwise similar to ${\mu^i}_j$, proceed as
follows. Following Seiberg we write the dual superpotential,
including a quark mass matrix ${\mu^i}_j$ as:
\label\dualW
\eq
{\widetilde W}\left(\widetilde M,\mu,\Omega, \widetilde S\right)
=\rho\tr\mu\widetilde M+\tr\Omega\widetilde M+{\lambda\over 3}
\tr\mu^3-N_c\left( \det\Omega \; e^{8\pi^2\widetilde S}\right)^{1/N_c} ,
\eeq
where $\rho$ is an  undetermined constant scale,
Here ${\Omega^i}_j\equiv \Avg{{q}^i\ol{q}_j}$, and we have added
the same cubic term in $\mu$ as used in the model of the previous
sections. Extremizing this superpotential with respect to
$\widetilde M$ implies ${\Omega^i}_j=-\rho{\mu^i}_j$, and
solving the field equation for ${\mu^i}_j$ in the result
 gives ${\mu^i}_j=\left((-)^{N_f}\lambda^{N_c}(\rho)^{-N_f}
e^{-8\pi^2\widetilde S}\right)^{1/(N_f-3N_c)}\delta^i_j$. This gives the
dual superpotential:
\label\dualWt
\eq
{\widetilde W}(\widetilde S)=\tilde k\left(\left(\lambda\rho^{-3}\right)^{N_f}
e^{-24\pi^2\widetilde{S}}\right)^{1/(N_f-3N_c)}  ,
\eeq
where $\tilde{k}=(-)^{-3N_f/(N_f-3N_c)}\left(N_f-3N_c\right)/3$.
This gives $\widetilde{W}$ proportional to a negative exponential
of $\widetilde S$, as claimed. The problem is to see the consistency
of this result with $W$ being proportional to a positive exponential
of $S$.

The resolution to this puzzle hinges on the connection
between the gauge couplings in the two models.
This is given by the duality relation for the RG
invariant scales of the two theories, which was
proposed for the case $N_f < 3 N_c$ in ref.~\seiberg,
and which we extend here to general $N_f$ and $N_c$.
Equating the {\it v.e.v.} found for ${\mu^i}_j(\widetilde S)$
in the dual theory with that found for ${\mu^i}_j(S)$
in the original model implies the relation:
\label\funnydual
\eq
\Lambda^{3N_c-N_f}\widetilde{\Lambda}^{2N_f-3N_c}=
(-)^{N_f-N_c}\, \rho^{N_f}.
\eeq
This duality relation in particular
implies that $\Re \widetilde S \to - \infty$
when $\Re S \to + \infty$.
Since $\widetilde W$ involves positive powers of
$e^{- \widetilde S}$, this is consistent with our
finding that $W$ involves positive powers of $e^{+S}$.

\ref\dbrana{K. Hori, H. Ooguri and Y. Oz, hep-th/9706082;
E. Witten, hep-th/9606109;
A. Brandhuber, N. Itzhaki, V. Kaplunovsky,
J. Sonnenschein and S. Yankielowicz, hep-th/9706127.}

We conclude, therefore, that string theory may well solve its
runaway dilaton problem by producing a low energy spectrum
which is described by a nonasymptotically-free effective
theory. If so, then run-of-the-mill mechanisms for supersymmetry
breaking --- like gaugino condensation --- can generate
superpotentials for which the dilaton {\it v.e.v.} is fixed
and cannot run off to weak coupling ($S \to \infty$). Thus,
nonasymptotically free theories bear more detailed study
to see whether this kind of mechanism can avoid some of
the other problems (such as fine-tuning or cosmological
constant problems) to which string theories seem prone.
It would be interesting to find a connection between the
field theoretical ideas discussed in this paper
and the recent developments understanding $N=1$ supersymmetric theories
from the $D$-branes structure of $M$-theory \dbrana.

\bigskip
\centerline{\bf Acknowledgments}
\bigskip

We thank P. Argyres,
L. Ib\'a\~nez, E. Poppitz, M. Quir\'os
and G. Veneziano for useful conversations,
and R. Myers for making the remark which
initiated this line of inquiry.
Our research was partially funded
by the N.S.E.R.C.\ of Canada and
the UNAM/McGill collaboration agreement.

\listrefs

\bye